\documentclass[prd,aps,preprintnumbers,floatfix,showpacs,preprint,tightenlines]{revtex4}
\usepackage{epsfig}
\usepackage{graphics}
\usepackage{latexsym}
\usepackage{amsmath}
\usepackage{amssymb}
\usepackage{rotating}

\begin{document}
\preprint{ \vbox{ \hbox{JLAB-THY-06-545}  \hbox{ADP-06-06/T637} }}

\title {Phase transition from hadronic matter to quark matter}

\author{P. Wang$^{ac}$}
\author{A. W. Thomas$^b$}
\author{A. G. Williams$^c$}

\affiliation{ $^a$Physics Department, North Carolina State
University, Raleigh, NC 27695, USA}

\affiliation{ $^b$Jefferson Laboratory, 12000 Jefferson Ave.,
Newport News, VA 23606, USA}

\affiliation{ $^c$Special Research Center for the Subatomic
Structure of Matter (CSSM) and Department of Physics, University of
Adelaide, Adelaide, SA 5005, Australia}

\begin{abstract}
We study the phase transition from two-flavor nuclear matter
to quark matter. A mean field model, constructed at the quark level,
is used to give the equation of state for
nuclear matter, while the equation of state for color
superconducting quark matter is calculated within the NJL model. It
is found that at low temperature, the phase transition from nuclear
to color superconducting quark matter will take place when the
density is of order 2.5$\rho_0$ - 5$\rho_0$. At zero density, the
quark phase will appear when the temperature is larger than about
148 MeV. Within the mean field treatment, the phase transition from
nuclear matter to quark matter is
always first order, whereas the transition between color
superconducting quark matter and normal quark matter is second
order.
\end{abstract}


\maketitle

\section{Introduction}

Hadronic matter is expected to undergo a phase transition to quark
matter at high temperature and/or high baryon density. Quark matter
may exist in the core of neutron star where the density is high or
it may be produced in the laboratory in heavy ion collisions.
In quark matter, the presence of a weak attraction
between the quarks will result in the formation of a condensate of quark
pairs. Pairs of quarks cannot be color singlets, and in QCD with two
flavors of massless quarks, the Cooper pairs favor the formation of a color
$\bar{3}$, flavor singlet condensate. The color
superconductivity of quark matter has been discussed in many
places in the literature (see for example \cite{Sara} - \cite{Agasian} and
references therein).

The phase diagram for chiral symmetry
restoration, including possible color superconductivity, for
quark matter containing just u and d
quarks, was discussed in Refs.~\cite{Berges,Schwarz}, where a rich
phase structure was obtained. It was shown that at high temperature,
chiral symmetry was restored via a second-order phase transition. At
low temperature, the phase transition was first order and at this
time, a new phase with a condensate of Cooper pairs appeared.
However, we should notice that this phase transition is not a
transition from hadronic matter to quark matter but a transition
from chiral symmetry breaking quark matter to chiral symmetry
restored quark matter. We know that in the real world, there is
no quark matter at low density and temperature, rather the chiral symmetry
breaking quark matter should be replaced by hadronic matter.
Therefore, when studying the phase diagram, the equations of state
(EOS) of hadronic matter are needed.

To study the properties of hadronic matter we need phenomenological
models, since QCD cannot yet be used directly. The symmetries of QCD
can be used to constrain the hadronic interactions and models based
on $SU(2)_{L}\times SU(2)_{R}$ symmetry and scale invariance have
been proposed. These effective models have been widely used to
investigate nuclear matter and finite nuclei, both at zero and at
finite temperature \cite{Furnstahl}-\cite{Zhang2}. Papazoglou $et$
$al.$ extended the chiral effective models to $SU(3)_{L}\times
SU(3)_{R}$, including the baryon
octet\cite{Papazoglou1,Papazoglou2}. As well as models based on
hadronic degrees of freedom, there are others based on
quark degrees of freedom, such as the quark meson coupling model
\cite{Guichon,Kazuo}, the cloudy bag model \cite{Thomas}, the
Nambu-Jona-Lasinio (NJL) model \cite{Bentz} and the quark mean field
model \cite{Toki}, $etc.$.
Several years ago, a chiral $SU(3)$ quark mean field model based on
quark degrees of freedom was proposed by Wang $et$ $al$.
\cite{Wang3,Wang4}. In this model, quarks are confined in the
baryons by an effective potential. The quark-meson interaction and
meson self-interaction are based on $SU(3)$ chiral symmetry. Through
the mechanism of spontaneous symmetry breaking the resulting
constituent quarks and mesons (except for the pseudoscalars) obtain
masses. The introduction of an explicit symmetry breaking term in
the meson self-interaction generates the masses of the pseudoscalar
mesons, which satisfy the relevant PCAC relations. The explicit
symmetry breaking term in the quark-meson interaction gives
reasonable hyperon potentials in hadronic matter. This chiral
$SU(3)$ quark mean field model has been applied to investigate
nuclear matter \cite{Wang2}, strange hadronic matter \cite{Wang3},
finite nuclei, hypernuclei \cite{Wang4}, and quark matter
\cite{Wang5}. Recently, we improved the chiral $SU(3)$ quark mean
field model by using a microscopically justified treatment of the
center of mass correction~\cite{Wangcssm1}.
This new treatment has been applied to the study of the
liquid-gas phase transition of asymmetric nuclear matter as well as strange
hadronic matter \cite{Wangcssm2,Wangcssm3}. By and large the results
are in reasonable agreement with existing experimental data.

As we know, the phase transition between hadronic matter and quark
matter can only occur at high density or temperature. As the
density grows, one expects that the nucleons will begin to overlap. This
is expected to be a source of short distance repulsion and correlations.
Such effects are also encountered in atomic physics and a
method of estimating the effect was introduced in
Ref.~\cite{ATOMIC}. This was later applied in the context of
dense nuclear matter by~\cite{Panda} and we also use
this method to estimate the importance of the effect.
As we will show later, this procedure generates
a repulsive interaction between the nucleons and increases their
chemical potentials in nuclear matter.

In this paper, we study the phase transition from nuclear matter to
quark matter. The effect of nucleon finite size is considered in
nuclear matter and color superconductivity is discussed in quark
matter. Since this is the first time we study the volume effect on
the phase transition in our model, we restrict our investigation
to just two flavors,
without including strange quarks or hyperons. The paper is
organized in the following way. Section II sets out the formalism of
the NJL model which is applied to describe quark matter. In section
III we introduce the $SU(3)$ quark mean field model which is used to
describe nuclear matter -- although in this work only the
non-strange sector of the model is required. The effect of finite
nucleon size is discussed in this section. The numerical results are
presented in section IV. Section V is the summary.

\section{NJL model}

In order to study the color superconductivity of quark
matter, we choose the NJL model, within which it is simple to incorporate
the diquark condensate
in the Lagrangian. The Lagrangian density for the $SU(2)$ NJL model
\cite{Nambu} can be written as
\begin{equation}
\pounds_1=\bar{\psi}_\alpha^i\left( i\gamma
^\mu\partial_\mu-M\right)
\psi_i^\alpha+G_1\left[\bar{\psi}_\alpha^i\psi_i^\alpha
\bar{\psi}_\beta^j\psi_j^\beta-\bar{\psi}_\alpha^i\gamma_5\left(\tau_a\right)_i^j
\psi_j^\alpha\bar{\psi}_\beta^k\gamma_5\left(\tau_a\right)_k^l\psi_l^\beta\right]-B^\prime,
\end{equation}
where $M$ is the current mass matrix of the u and d quarks, $\tau_a$
($a=1-3$) are the Pauli matrices. $B^\prime$ is the parameter which
will give a reasonable bag constant, $B$, in quark matter. In order
to describe the diquark condensate, the interaction of the color
$\bar{3}$ fermion bilinears should be added. This four-fermion
interaction can be written as
\begin{equation}
\pounds_2=G_2\left[\left(\bar{\psi}^T\right)^i_\beta C\gamma^5\varepsilon^{\alpha\beta\gamma}
\varepsilon_{ij}\bar{\psi}^j_\gamma\right]\left(\psi^T\right)^k_\rho C\gamma^5\varepsilon^{\alpha
\rho\sigma}\varepsilon_{kl}\psi^l_\sigma,
\end{equation}
where $C$ denotes the charge conjugation matrix. In the above
equations, $G_1$ and $G_2$ are the coupling constants.

From the above Lagrangian, the thermodynamic potential $\Omega$,
which determines the field equations can be obtained. It is
convenience to introduce bosonic collective fields into the
functional integral to make the four fermion interaction into
a fermion-boson coupling. We follow Ref.~\cite{Berges} to employ a
Hubbard-Stratonovich transformation, in which the collective fields
are introduced into the functional integral by inserting the
following identities
\begin{equation}
1=N_1\int
D\phi_q\exp\left\{i\int\frac{d^4p}{(2\pi)^4}\left(\frac{\phi_q}{2}-
G_1\bar{\psi}_\alpha^i\psi_i^\alpha\right)\frac{1}{G_1}\left(\frac{\phi_q}{2}
-G_1\bar{\psi}_\alpha^i\psi_i^\alpha\right)\right\},
\end{equation}
\begin{eqnarray}
1&=&N_2\int D\Delta_q^*D\Delta_q\exp\left\{i\int\frac{d^4p}{(2\pi)^4}\left(
\frac{\Delta_q^{*\alpha}}{2}-
G_2\left[\left(\bar{\psi}^T\right)^i_\beta C\gamma^5\varepsilon^{\alpha\beta\gamma}
\varepsilon_{ij}\bar{\psi}^j_\gamma\right]\right) \right .
\nonumber \\
&& \left .\times\frac{1}{G_2}
\left(\frac{\Delta_q^\alpha}{2}-G_2\left(\psi^T\right)^i_\beta C\gamma^5
\varepsilon^{\alpha\beta\gamma}
\varepsilon_{ij}\psi^j_\gamma\right)\right\}.
\end{eqnarray}
The quadratic terms in equations (3) and (4) cancel the original
four-fermion interaction. As a result, the four-fermion interaction
is cast into a Yukawa interaction between fermions and collective
fields and a mass term for the collective fields. The fermionic
fields then appear only quadratically and can be integrated out
exactly, leaving an effective action for the bosonic collective
fields alone. Having done the Gaussian integral for the fermions,
the thermodynamic potential $\Omega_q$ reads
\begin{eqnarray}
\Omega_q[\phi_q, \Delta_q; T, \mu]&=&\frac{1}{4G_1}\phi_q^2
+\frac{1}{4G_2}\Delta_q^2 + B^\prime
\nonumber \\
&&+\frac{i}{2}\ln\det\left(
\begin{array}{lr}
\gamma^\mu p_\mu+\gamma^0\mu_q-(m_q+\phi_q) & {\bf\Delta_q}~~~~~~~~~~ \\
~~~~~~~~~~{\bf\Delta_q^*} & [\gamma^\mu p_\mu+\gamma^0\mu_0-(m_q+\phi_q)]^T  \\
\end{array}
\right). \nonumber \\
\end{eqnarray}
Disregarding the $\gamma$ matrix, the matrix of the last term of the
above equation is $12\times 12$. ${\bf\Delta_q}$ is a $6\times 6$
matrix whose matrix elements are
$\left({\bf\Delta_q}\right)^{i\alpha,
j\beta}=C\gamma^5\Delta_q\varepsilon_{ij}
\varepsilon^{3\alpha\beta}$, where $i$, $j$ are flavor indices and
$\alpha, \beta$ are color indices.

The thermodynamic potential can be obtained by using imaginary time
Green function method, with $p_0=(2n+1)\pi/\beta$ ($\beta$=1/T).
After performing the sums of the Matsubara frequences, the
thermodynamic potential $\Omega$ is written as
\begin{eqnarray}
\Omega&=&\frac{1}{4G_1}\phi^2+\frac{1}{4G_2}\Delta^2
-\sum_{\tau=u,d}\int\frac{p^2dp}{\pi^2}\left\{E_\tau
+T\ln\left(1+e^{-\beta\left(E_\tau-\mu_\tau\right)}\right) \right . \nonumber
\\ && \left .+T\ln\left(1+e^{-\beta\left(E_\tau+\mu_\tau\right)}\right)
+\sqrt{\zeta_{\tau+}^2+\Delta^2}
+\theta\left(\zeta_{\tau-}\right)\sqrt{\zeta_{\tau-}^2+\Delta^2}
\right .\nonumber \\
&& \left .+2T\ln\left(1+e^{-\beta\sqrt{\zeta_{\tau+}^2+\Delta^2}}\right)
+2T\ln\left(1+e^{-\beta\theta\left(\zeta_{\tau-}\right)
\sqrt{\zeta_{\tau-}^2+\Delta^2}}\right)\right\}+B^\prime,
\end{eqnarray}
where $\zeta_{\tau\pm}=E_\tau\pm\mu_\tau$,
$E_\tau=\sqrt{m_\tau^{*2}+p^2}$ with $m_\tau^*=m_{\tau 0}+\phi
$ and $\theta(x)$ is a step function defined as
\begin{eqnarray}
\theta(x)=\biggl\{
\begin{array}{lr}
1 & ~~~~~~ x> 0 \\
-1 & ~~~~~~ x<0 \\
\end{array}
\end{eqnarray}
The quark condensate $\langle\bar{\psi}\psi\rangle$ is related to $\phi$ via
$\phi=-2G_1\langle\bar{\psi}\psi\rangle$.

In the four fermion interaction, form factors are not included
\cite{Berges}. Following Ref.~\cite{Schwarz}, we introduce a cutoff,
$\Lambda_q$ to regulate the above integral. After minimizing the
thermodynamic potential with respect to $\phi$ and $\Delta$, the
following gap equations can be obtained.
\begin{eqnarray}
m_\tau^*&=& m_{\tau 0}+4G_1\int\frac{m_\tau^*p^2dp}{\pi^2E_\tau}\left\{1-
n_F(E_\tau)-\bar{n}_F(E_\tau)+\frac{\zeta_{\tau+}}{\sqrt{\zeta_{\tau+}^2
+\Delta_\tau^2}}\left[1-2n_d(\zeta_{\tau+})\right] \right. \nonumber \\
&&\left.+\frac{\theta(\zeta_{\tau-})\zeta_{\tau-}}{\sqrt{\zeta_{\tau-}^2
+\Delta_\tau^2}}\left[1-2\bar{n}_d(\zeta_{\tau-})\right]\right\},
\end{eqnarray}
\begin{eqnarray}
\Delta_\tau&=& 4G_2\int\frac{p^2dp}{\pi^2}\left\{
\frac{\Delta_\tau}{\sqrt{\zeta_{\tau+}^2+\Delta_\tau^2}}
\left[1-2n_d(\zeta_{\tau+})\right]
+\frac{\theta(\zeta_{\tau-})\Delta_\tau}{\sqrt{\zeta_{\tau-}^2+\Delta_\tau^2}}
\left[1-2\bar{n}_d(\zeta_{\tau-})\right]\right\},
\end{eqnarray}
where $n_F(E_\tau)$ and $\bar{n}_F(E_\tau)$ are the quark and anti-quark
distributions, respectively, expressed as
\begin{equation}
n_F(E_\tau)=\left\{\exp\left[\left(E_\tau-\mu_\tau\right)/T\right]+1\right\}^{-1},
\end{equation}
\begin{equation}
\bar{n}_F(E_\tau)=\left\{\exp\left[\left(E_\tau+\mu_\tau\right)
/T\right]+1\right\}^{-1}.
\end{equation}
The function $n_d(\zeta_{\tau+})$ and $\bar{n}_d(\zeta_{\tau-})$ are defined as
\begin{equation}
n_d(\zeta_{\tau+})=\left\{\exp\left[\sqrt{\zeta_{\tau+}^2+\Delta_\tau^2}/T\right]+1
\right\}^{-1},
\end{equation}
\begin{equation}
\bar{n}_d(\zeta_{\tau-})=\left\{\exp\left[\theta(\zeta_{\tau-})\sqrt{\zeta_{\tau-}^2
+\Delta_\tau^2}/T\right]+1\right\}^{-1}.
\end{equation}
From the thermodynamic potential $\Omega$, the pressure and quark
density can be obtained as $p_Q=-\Omega$ and
$\rho_\tau=\partial\Omega/\partial\mu_\tau$. One can define baryon
density as $\rho_N=\frac{1}{3}\left(\rho_u+\rho_d\right)$. In the
chiral limit, the current quark mass $m_{u0}$=$m_{d0}=0$. In this
case, equation (8) has a trivial solution $m^*=0$. The gap equation
for $\Delta_q$ also has a trivial solution $\Delta=0$. Therefore,
for the massless quark, at some range of chemical potential or
baryon density, there exist four sets of solutions: A. $m^*\neq 0$,
$\Delta\neq 0$; B. $m^*\neq 0$, $\Delta=0$; C. $m^*=0$, $\Delta\neq
0$; D. $m^*=0$, $\Delta=0$. The system favors being in the phase
with lowest thermodynamic potential. By comparing the
thermodynamical potentials, one finds that the system will be in
the phase with $m^*\neq 0$, $\Delta=0$ at low chemical potential
(temperature) and in the phase with $m^*=0$ at large chemical
potential (temperature). However, at low chemical potential
(temperature), the system cannot be in the quark phase but in the
hadronic phase. Therefore, we need to replace the EOS for chiral
symmetry breaking quark matter by that for hadronic matter.

\section{Nuclear matter}

For hadronic matter it is possible to work from the quark level, as in
the quark meson coupling model or even in the NJL model. However, in the
NJL model it is essential to incorporate the effect of confinement
(for example through proper time regularization) in order to avoid a
chiral collapse~\cite{Bentz}. In the present work we adopt the more
common hybrid approach and use the $SU(3)$ quark mean field model in
the hadronic phase.
In that model the total effective Lagrangian is written:
\begin{eqnarray}
{\cal L}_{{\rm eff}} \, = \, {\cal L}_{q0} \, + \, {\cal L}_{qM}
\, + \,
{\cal L}_{\Sigma\Sigma} \,+\, {\cal L}_{VV} \,+\, {\cal L}_{\chi SB}\,
+ \, {\cal L}_{\Delta m_s} \, + \, {\cal L}_{h}, + \, {\cal L}_{c},
\end{eqnarray}
where ${\cal L}_{q0} =\bar q \, i\gamma^\mu \partial_\mu \, q$ is the
free part for massless quarks. The quark-meson interaction
${\cal L}_{qM}$ can be written in a chiral $SU(3)$ invariant way as
\begin{eqnarray}
{\cal L}_{qM}=g_s\left(\bar{\Psi}_LM\Psi_R+\bar{\Psi}_RM^+\Psi_L\right)
-g_v\left(\bar{\Psi}_L\gamma^\mu l_\mu\Psi_L+\bar{\Psi}_R\gamma^\mu
r_\mu\Psi_R\right)~~~~~~~~~~~~~~~~~~~~~~~  \nonumber \\
=\frac{g_s}{\sqrt{2}}\bar{\Psi}\left(\sum_{a=0}^8 s_a\lambda_a
+ i \gamma^5 \sum_{a=0}^8 p_a\lambda_a
\right)\Psi -\frac{g_v}{2\sqrt{2}}
\bar{\Psi}\left( \gamma^\mu \sum_{a=0}^8
 v_\mu^a\lambda_a
- \gamma^\mu\gamma^5 \sum_{a=0}^8
a_\mu^a\lambda_a\right)\Psi.
\end{eqnarray}
In the mean field approximation, the chiral-invariant scalar meson
${\cal L}_{\Sigma\Sigma}$ and vector meson ${\cal L}_{VV}$
self-interaction terms are written as
\begin{eqnarray}
{\cal L}_{\Sigma\Sigma} &=& -\frac{1}{2} \, k_0\chi^2
\left(\sigma^2+\zeta^2\right)+k_1 \left(\sigma^2+\zeta^2\right)^2
+k_2\left(\frac{\sigma^4}2 +\zeta^4\right)+k_3\chi\sigma^2\zeta
\nonumber \\ \label{scalar}
&&-k_4\chi^4-\frac14\chi^4 \ln \frac{\chi^4}{\chi_0^4} +
\frac{\delta}
3\chi^4 \ln \frac{\sigma^2\zeta}{\sigma_0^2\zeta_0}, \\
{\cal L}_{VV}&=&\frac{1}{2} \, \frac{\chi^2}{\chi_0^2} \left(
m_\omega^2\omega^2+m_\rho^2\rho^2+m_\phi^2\phi^2\right)+g_4
\left(\omega^4+6\omega^2\rho^2+\rho^4+2\phi^4\right), \label{vector}
\end{eqnarray}
where $\delta = 6/33$; $\sigma_0$, $\zeta_0$ and $\chi_0$ are the
vacuum expectation values of the corresponding mean fields
$\sigma$, $\zeta$ and $\chi$.

The Lagrangian ${\cal L}_{\chi SB}$ generates
nonvanishing masses for the pseudoscalar mesons
\begin{equation}\label{L_SB}
{\cal L}_{\chi SB}=\frac{\chi^2}{\chi_0^2}\left[m_\pi^2F_\pi\sigma +
\left(
\sqrt{2} \, m_K^2F_K-\frac{m_\pi^2}{\sqrt{2}} F_\pi\right)\zeta\right],
\end{equation}
leading to a nonvanishing divergence of the axial currents which in
turn satisfy the partial conserved axial-vector current (PCAC)
relations for $\pi$ and $K$ mesons. Pseudoscalar,
scalar mesons and also the dilaton field $\chi$ obtain mass terms by
spontaneous breaking of chiral symmetry in the Lagrangian
(\ref{scalar}). The masses of $u$, $d$ and $s$ quarks are generated by
the vacuum expectation values of the two scalar mesons $\sigma$ and
$\zeta$.

In the quark mean field model, quarks are confined in baryons by the
Lagrangian ${\cal L}_c=-\bar{\Psi} \, \chi_c \, \Psi$. $\chi_c(r)$
is a confinement potential, i.e. a static potential providing the
confinement of quarks by meson mean-field configurations. In the
numerical calculations, we choose $\chi_{c}(r)= \frac14 k_cr^2$,
where $k_{c} = 1$ GeV fm$^{-2}$. The Dirac equation for a quark
field $\Psi_{ij}$ under the additional influence of the meson mean
fields is given by
\begin{equation}
\left[-i\vec{\alpha}\cdot\vec{\nabla}+\beta \chi_c(r)+\beta
m_i^*\right] \Psi_{ij}=e_i^*\Psi_{ij}, \label{Dirac}
\end{equation}
where $\vec{\alpha} = \gamma^0 \vec{\gamma}$, $\beta = \gamma^0$,
the subscripts $i$ and $j$ denote the quark $i$ ($i=u, d, s$) in a
baryon of type $j$ ($j=N, \Lambda, \Sigma, \Xi$). The quark mass
$m_i^*$ and energy $e_i^*$ are defined as
\begin{equation}
m_i^*=-g_\sigma^i\sigma - g_\zeta^i\zeta+m_{i0}
\end{equation}
and
\begin{equation}
e_i^*=e_i-g_\omega^i\omega-g_\phi^i\phi \,,
\end{equation}
where $e_i$ is the energy of the quark under the influence of the
meson mean fields. The effective baryon mass can be written as
\begin{eqnarray}
M_j^*=\sum_in_{ij}e_i^*-E_j^0 \label{linear}\,,
\end{eqnarray}
where $n_{ij}$ is the number of quarks with flavor $``i"$ in a
baryon $j$ and $E_j^0$ is adjusted to give a best fit to the free
baryon mass.

Based on the previously defined quark mean field model
the Lagrangian density for nuclear matter is written as
\begin{eqnarray}
{\cal L}&=&\bar{\psi}(i\gamma^\mu\partial_\mu-M_N^*)\psi
+\frac12\partial_\mu\sigma\partial^\mu\sigma+\frac12
\partial_\mu\zeta\partial^\mu\zeta+\frac12\partial_\mu
\chi\partial^\mu\chi-\frac14F_{\mu\nu}F^{\mu\nu} -
\frac14 \rho_{\mu\nu}\rho^{\mu\nu}\nonumber \\
&&-g_\omega\bar{\psi}\gamma_\mu\psi\omega^\mu -g_\rho\bar{\psi}
_B\gamma_\mu\tau_3\psi\rho^\mu +{\cal L}_M, \label{bmeson}
\end{eqnarray}
where
\begin{equation}
F_{\mu\nu}=\partial_\mu\omega_\nu-\partial_\nu\omega_\mu
\hspace*{.5cm} \mbox{and} \hspace*{.5cm}
\rho_{\mu\nu}=\partial_\mu\rho_\nu-\partial_\nu\rho_\mu.
\end{equation}
The term ${\cal L}_M$ represents the interaction between mesons which
includes the scalar meson self-interaction ${\cal L}_{\Sigma\Sigma}$,
the vector meson self-interaction
${\cal L}_{VV}$ and the explicit chiral symmetry breaking term
${\cal L}_{\chi SB}$, all defined previously.
The Lagrangian includes the scalar mesons
$\sigma$, $\zeta$ and $\chi$, and the vector mesons $\omega$ and $\rho$.
The interactions between quarks and scalar mesons result in the effective
nucleon mass $M_N^*$, The interactions between quarks and
vector mesons generate the nucleon-vector meson interaction terms of
equation (\ref{bmeson}).
The corresponding vector coupling constants $g_\omega$ and $g_\rho$ are
baryon dependent and satisfy the $SU(3)$ relationship:
$g_\rho^p=-g_\rho^n=\frac13 g_\omega^p=\frac13 g_\omega^n$.

At finite temperature and density, the thermodynamic potential is
defined as
\begin{eqnarray}
\Omega = - \frac{k_{B}T}{(2\pi)^3}\sum_{N=p,n}
\int_0^\infty d^3\overrightarrow{k}\biggl\{{\rm ln}
\left( 1+e^{-(E_N^{\ast}(k) - \nu _N)/k_{B}T}\right)
+ {\rm ln}\left( 1+e^{-(E_N^{\ast }(k)+\nu_N)/k_{B}T}
\right) \biggr\} -{\cal L}_{M},
\end{eqnarray}
where $E_N^{\ast }(k)=\sqrt{M_N^{\ast 2}+\overrightarrow{k}^2}$.
The quantity $\nu _N$ is related to the usual chemical potential,
$\mu _N$, by $\nu _N =\mu _N-g_{\omega }^N\omega -g_{\rho }^N\rho$.
The energy per unit volume and the pressure of the system are respectively
$\varepsilon =\Omega -\frac1T
\frac{\partial\Omega}{\partial T}+\nu _N\rho_N$ and $p_H=-\Omega $,
where $\rho_N$ is the baryon density.

The mean field equation for meson $\phi _{i}$ is obtained by the
formula $\partial \Omega/\partial \phi_i=0$. For example,
the equations for $\sigma$, $\zeta$ are deduced as:
\begin{eqnarray}\label{eq_sigma}
k_{0}\chi ^{2}\sigma
-4k_{1}\left( \sigma ^{2}+\zeta ^{2}\right) \sigma -2k_{2}\sigma
^{3}-2k_{3}\chi \sigma \zeta -\frac{2\delta }{3\sigma }\chi ^{4}
+\frac{\chi^{2}}{\chi _{0}^{2}}m_{\pi }^{2}F_{\pi }  \nonumber \\
-\left( \frac{\chi }{\chi _{0}}\right) ^{2}m_{\omega }\omega ^{2}\frac{
\partial m_{\omega }}{\partial \sigma }
-\left( \frac{\chi }{\chi _{0}}\right) ^{2}m_{\rho }\rho ^{2}\frac{
\partial m_{\rho }}{\partial \sigma }
+\frac{\partial M_N^{\ast }}{\partial \sigma } \langle\bar{\psi}\psi\rangle=0,
\end{eqnarray}
\begin{eqnarray}\label{eq_zeta}
k_{0}\chi ^{2}\zeta -4k_{1}\left(\sigma ^{2}+\zeta ^{2}\right)
\zeta -4k_{2}\zeta ^{3}-k_{3}\chi \sigma ^{2} -
\frac{\delta }{3\zeta }\chi ^{4}+\frac{\chi ^{2}}{\chi _{0}^{2}}
\left( \sqrt{2}m_{k}^{2}F_{k}-\frac{1}{\sqrt{2}}m_{\pi }^{2}F_{\pi }
\right)=0,
\end{eqnarray}
where
\begin{equation}
\langle\bar{\psi}\psi\rangle=\frac{1}{\pi ^{2}}\int_{0}^{\infty}
dk \frac{k^{2}M_{N}^{\ast }}{E^*(k)}
\left[n_n(k)+\bar{n}_n(k)+n_p(k)+\bar{n}_p(k)\right].
\end{equation}
In the above equation, $n_N(k)$ and $\bar{n}_N(k)$ are the nucleon
and antinucleon distributions, respectively, expressed as
\begin{equation}
n_N(k)=\frac{1}{\exp\left[\left(E^*(k)-\nu_N\right)/k_B T\right]+1}
\end{equation}
and
\begin{equation}
\bar{n}_N(k)=\frac{1}{\exp\left[\left(E^*(k)+\nu_N\right)/k_B
T\right]+1} ~~~~(N=n,p).
\end{equation}
In the vacuum, the thermodynamic potential can have another solution where
$\sigma=0$, $\zeta=0$ and $\chi=0$. The energy difference of these
two kinds of vacua is
\begin{equation}
\Delta E = {\cal L}_M(\sigma=0,\zeta=0,\chi=0)-{\cal
L}_M(\sigma=\sigma_0,\zeta=\zeta_0,\chi=\chi_0).
\end{equation}
It is interesting that the calculated value $\Delta E$ is close to
the bag constant of quark matter. The numerical values of $\Delta E$
are listed in Table II.

In the same way, the equations for the vector mesons, $\omega$ and
$\rho$, can be obtained:
\begin{equation}
\frac{\chi^2}{\chi_0^2}m_\omega^2\omega+4g_4\omega^3+12g_4\omega\rho^2
=g_\omega^N(\rho_p+\rho_n),
\end{equation}
\begin{equation}
\frac{\chi^2}{\chi_0^2}m_\rho^2\rho+4g_4\rho^3+12g_4\omega^2\rho
=\frac13 g_\omega^N(\rho_p-\rho_n),
\end{equation}
where $\rho_p$ and $\rho_n$ are the proton and neutron densities,
expressed as
\begin{equation}
\rho_N=\frac{1}{\pi ^{2}}\int_{0}^{\infty}
dk k^2\left[n_q(k)-\bar{n}_q(k)\right] ~~~~ (N=p,n).
\end{equation}

In nuclear matter, when the density is high, the volume of the
nucleon is important. The nucleon cannot be treated as a point-like
particle. The excluded volume effect has been considered in
Ref.~\cite{Panda}. The key point is that the pressure of a system with
excluded volume is the same as that of a point particle
system, $p_H(T,\tilde{\mu})$, if the chemical potentials of these two
systems have the following relationship
\begin{equation}
\mu=\tilde{\mu}+v_0p_B(T,\tilde{\mu}),
\end{equation}
where $v_0$ is the volume of a nucleon. $p_B(T,\tilde{\mu})$ is the
pressure generated by baryons, i.e. $p_B(T,\tilde{\mu})=p_H-{\cal
L}_M$. The baryon density, entropy density and the energy density of
the system (without meson contribution) are given by the usual
thermodynamic expressions. We find the following relations:
\begin{equation}
\rho_B=\left( \frac{\partial  p_B}{\partial \mu}\right)_T =
\left( \frac{\partial  p_B}{\partial \tilde{\mu}}\right)_T
\frac{\partial \tilde{\mu}}{\partial \mu}=
\frac{\tilde{\rho}_B}{1+v_0 \tilde{\rho}_B},
\end{equation}
\begin{equation}
S_B=\left(\frac{\partial p_B}{\partial T}\right)_\mu
=\left(\frac{\partial p_B}{\partial T}\right)_{\tilde{\mu}} +\left(
\frac{\partial  p_B}{\partial \tilde{\mu}}\right)_T
\left(\frac{\partial \tilde{\mu}}{\partial T}\right)_\mu
=\frac{\tilde{S}_B}{1+v_0\tilde{\rho}_B},
\end{equation}
\begin{equation}
\varepsilon_B=TS_B-p_B+\mu\rho_B=\frac{\tilde{\varepsilon}_B}{1+v_0
\tilde{\rho}_B}.
\end{equation}
When the effect of the finite volume of the nucleon is
included, the scalar density,
$\langle\bar{\psi}\psi\rangle$, in equation (26) becomes
\begin{equation}
\langle\bar{\psi}\psi\rangle=\frac{1}{\pi^{2}(1+v_0 \tilde{\rho}_B)}\int_{0}^{\infty}
dk \frac{k^{2}M_{N}^{\ast }}{E^*(k)}
\left[n_n(k)+\bar{n}_n(k)+n_p(k)+\bar{n}_p(k)\right].
\end{equation}
The phase transition from hadronic matter to quark matter is
determined by the following conditions:
\begin{equation}
p_H=p_Q, ~~~ \mu_p=2\mu_u+\mu_d, ~~~ \mu_n=\mu_u+2\mu_d.
\end{equation}
We should mention that at finite temperature, gluons make an
important contribution to the pressure of quark matter.
The gluon contribution to the pressure of quark matter is
expressed as
\begin{equation}
p_g=\frac{8}{3\pi^2}\int_0^\infty dk \frac{k^3}{e^{k/k_B T}-1}.
\end{equation}

\section{Numerical results}

In the NJL model, $\Lambda_q$ is chosen to obtain the quark
condensate $\langle\bar{q}q\rangle=-250$ MeV.  $G_1$ is determined by the bold
quark mass in vacuum $m_q=350$ MeV. There is some uncertainty in the
coupling constant $G_2$. The instanton interaction suggests
$G_2/G_1=1/2$, while in Ref. \cite{Berges}, this ratio is 3/4. In our
calculations we choose $G_2/G_1=0.6$, which is close to that of Ref.
\cite{Schwarz}. The parameter $B^\prime$ is determined by choosing
the bag constant of quark matter to be 209MeV, i.e.
$p_Q(m_q^*=0,\mu_q=0)=-(209$MeV$)^4$. The numerical values of these
three parameters are: $\Lambda_q=634.7$ MeV, $G_1$=5.6 GeV$^{-2}$
and $(B^\prime)^{1/4}=605.7$ MeV. As for the nuclear matter phase, the
parameters in the Lagrangian, which are shown in Table I,
are determined by the saturation
properties and the meson masses in vacuum.
The corresponding nuclear matter properties and scalar meson masses
are listed in
Table II.

In Fig.~1 we plot the nonzero solutions for the effective quark mass and
energy gap versus baryon density at zero temperature. From the
figure, one can see that the effective quark mass decreases
continuously to zero with the increasing density. The energy gap
also changes continuously with the density. However, at low
temperature, the effective quark mass cannot decrease to zero
continuously, as shown in Fig.~1. The energy gap cannot change
continuously either -- there is a first order phase transition. We
have mentioned in section II that there are four kinds of solutions
to Eqs.~(8) and (9). At a given chemical potential the system
will be in the phase with the lowest thermodynamic potential
(highest pressure). In Fig.~2 we plot the pressure versus quark
chemical potential for the four kinds of solutions at zero
temperature. The solid, dashed, dotted and dash-dotted lines are for
$m^*=0$, $\Delta=0$ (phase D); $m^*=0$, $\Delta\neq 0$ (phase C);
$m^*\neq 0$, $\Delta\neq 0$ (phase A) and $m^*\neq 0$, $\Delta=0$
(phase B), respectively. At small chemical potential, the system
will be in the phase B: $\Delta=0$, $m^*\neq 0$. At large chemical
potential, the system will be in the phase C: $\Delta\neq 0$,
$m^*=0$. The phase transition takes place when the chemical
potential is about 342 MeV. Because this chemical potential is
smaller than the effective quark mass in phase B, the density of
this phase is zero. The effective mass and the energy gap versus
chemical potential are shown in Fig.~3. At the transition point,
$\mu_q=342$ MeV, the changes in these quantities are discontinuous.

With increasing temperature the energy gap will decrease. Fig.~4
shows the energy gap versus density at different temperatures. When
the temperature is larger than about 52 MeV the Cooper pairs cannot
form and the energy gap disappears. Fig.~5 shows the corresponding
phase diagram. The solid and dashed lines indicate first and second
order phase transitions, respectively. At low temperature, the
transition from phase B to phase C is of first order. The transition
from phase C to phase D is second order. This means that the two
$p-\mu$ curves, corresponding to phase C and D, turn into a single
curve at the transition point. We should mention that the result of
the second order phase transition is an artifact of the mean field
approximation. Early in 1974, Halperin, Lubensky and Ma have shown
that the superconducting to normal phase transition can not be the
second order \cite{Halperin}. In fact, this phase transition must be
of the first order when gauge field fluctuation is included
\cite{Giannakis,Giannakis2,Matsuura,Noronha}. At high temperature
the effective mass drops to zero continuously. The two $p-\mu$
curves of phase B and D turn into one curve smoothly and they do not
cross each other as at low temperature.

However, this phase diagram is for the transition from quark matter
to quark matter, not for the transition from nuclear matter to quark
matter. In nuclear matter the ``elementary particles" are protons
and neutrons. The equations of state for quark matter and nuclear
matter are different. The energy per baryon for nuclear matter is
shown in Fig.~6, where nucleon volume effects are considered. The
three curves correspond to the nucleon radius $R$ = 0, 0.6 fm and
0.75 fm, respectively. The inclusion of finite volume effects makes the
EOS for nuclear matter harder and generates a strong repulsive interaction
between nucleons.

The $p-\mu$ curves for nuclear matter and quark matter at zero
temperature are shown in Fig.~7. The three solid curves are for
nuclear matter and the two dashed lines for quark matter. If the
nucleons are treated as point particles, there is no phase
transition between nuclear matter and quark matter in this model.
The chemical potentials for the nucleons are always smaller than the
corresponding sums of quark potentials. When the volume effects are
included, the nucleon chemical potentials will be increased and
there is a first order phase transition between color
superconducting quark matter and nuclear matter. The $T-\mu$ phase
diagram for the transition between nuclear matter and quark matter is
shown in Fig.~8. The solid and dashed lines indicate first and
second order phase transitions, respectively. The letter H indicates the
hadronic phase of nuclear matter. When the chemical potential is zero, the
nuclear matter will turn into quark matter at temperature $T=148$
MeV. At low temperature, the phase transition from nuclear matter to
color superconducting quark matter will take place when the
nucleon chemical potential, $\mu_N$, is about 1.5 - 1.7 GeV.

Figs.~9 and 10 show the $T-\rho$ phase diagram for $R$ = 0.6 fm and
0.75 fm, respectively. Since the phase transition is of first order,
the densities of nuclear matter and quark matter are different.
There is a coexistence state, M, in the figures. For the case of $R$
= 0.6 fm, at zero temperature, when the nuclear density is about
0.76 fm$^{-3}$, quark matter starts to appear with density around
1.17 fm$^{-3}$. For the case $R$ = 0.75 fm, the corresponding
densities of nuclear matter and quark matter are 0.44 fm$^{-3}$ and
0.85 fm$^{-3}$. From our calculation, we conclude that at low
temperature, when the nuclear density is of order $2.5\rho_0$ -
$5\rho_0$, where $\rho_0$ is the saturation density of symmetric
nuclear matter, there will be a phase transition. Since the phase
transition between color superconducting quark matter and normal
quark matter is of second order, the density remains the same during
the phase transition.

In our calculation, most parameters are fixed by well known physical
quantities. Besides the nucleon radius $R$, there is only one
parameter, the bag constant, which will sensitively affect the
transition density and temperature. We discuss how the bag constant
affects the transition temperature at zero baryon density. From the
Fig.~11, one can see that with increasing temperature, the pressure
of nuclear matter at zero baryon density increases very slowly.
Until the temperature is around 180 MeV, the pressure remains low.
However, for quark matter, the pressure increases very fast when the
temperature is larger than about 90 MeV. This is because the quark
mass is much smaller than the nucleon mass. If the bag constant lies
in the range $180<B^{1/4}<230$ MeV, the transition temperature at
zero baryon density is in the range $130<T<170$ MeV. In other words,
if the transition temperature is about 150 MeV the bag constant will
be around $(210$ MeV$)^4$ which means that our choice of bag
constant is reasonable.

\section{Summary}

We have investigated the phase transition from nuclear matter to
quark matter. The $SU(3)$ quark mean field model, including nucleon
volume effects, has been applied to get the EOS of nuclear matter.
Color superconducting quark matter has also been studied, using the NJL
model. The parameters in the models are determined by well known
physical quantities. We now summarize the main results.

The phase transition between nuclear matter and quark matter is
of first order. There is no critical temperature above
which the phase transition is of second order. The so called
second order phase transition is a transition from chiral symmetry
breaking quark matter to chiral symmetry restored quark matter.
However, in the real world, at low density or temperature, there is
no chiral symmetry breaking quark matter, but rather one has nuclear
matter. Therefore, the phase transition from nuclear matter to quark
matter is always a first order transition, since the EOS of
nuclear and quark matter are different.

The effects of the finite volume of the nucleon are very important,
especially at
high density. The volume effects generate a strong repulsive
interaction between the nucleons and increase their chemical potentials.
The transition density is sensitive to the nucleon radius. In our
numerical calculations, we chose $R$ = 0.6 fm and 0.75 fm. At low
temperature, when the nuclear density is of order 2.5$\rho_0$ -
5$\rho_0$, the color superconducting quark phase will appear with
a density 0.4 fm$^{-3}$ larger than that of nuclear matter.

Besides the nucleon radius, the phase transition is also
sensitive to the bag constant $B$. If the bag constant is in the
range 180 MeV $<B^{1/4}<240$ MeV, the transition temperature will be
130 MeV $<T<$ 170 MeV. The bag constant $B$ can be explained as the
energy difference, $\Delta E$, between two kinds of vacua. In the
$SU(3)$ quark mean field model, the calculated $\Delta E$ is
reasonable, giving a transition temperature around 150 MeV.

At low temperature, when the density increases, the nuclear
matter will always turn into color superconducting quark matter.
When the temperature is larger than about 52 MeV, the superconductivity
phase disappears. The transition from normal quark matter to color
superconducting quark matter is of second order, while the
transition from nuclear matter to quark matter is of first
order.

In conclusion, we note that our discussions concerning the order of
the phase transitions investigated here have been based upon mean
field theory. In fact, for the transition from superconducting quark
matter to normal quark matter, with the gluon field fluctuation,
this transition must be of the first order. It clearly will be
important to explore such effects in the current model. In addition,
we have presented results without any consideration of strangeness.
As strange quarks may be expected to appear at perhaps 2.5 to 3
times normal nuclear matter density (for matter in beta equilibrium,
as in a neutron star), it will also be of interest to extend the
present model to include strange quarks.

\begin{center}
\textbf{Acknowledgements}
\end{center}

This work was supported by Australian Research Council and by DOE contract DOE-AC05-06OR23177,
under which Jefferson Science Associates operates Jefferson Lab.

 \vfill

\begin{table}
\caption{Parameters of the $SU(3)$ quark mean field model.}
\begin{center}
\begin{tabular}{||c|c|c|c|c|c|c|c|c||}
R (fm) & $k_0$ & $k_1$ & $k_2$ & $k_3$ & $k_4$ & $g_s$ & $g_v$
& $g_4$ \\
\hline
0 & 3.97 & 2.18 & -10.16 & -4.15 & -0.14 & 4.76 & 8.70 & 15.0\\
0.6 & 3.95 & 1.83 & -10.16 & -4.56 & -0.10 & 4.76 & 8.70 & 15.0\\
0.75 & 3.94 & 1.40 & -10.16 & -5.13 & -0.03 & 4.76 & 8.70 & 15.0\\
\end{tabular}
\end{center}
\end{table}

\begin{table}
\caption{Nuclear matter properties and scalar meson masses in
the $SU(3)$ quark mean field model.}
\begin{center}
\begin{tabular}{||c|c|c|c|c|c|c|c||}
R (fm) & $\rho_0$ (fm$^{-3}$) & E/A (MeV) & $M^*_N/M_N$ & K (MeV)&
$(\Delta E)^{1/4}$ (MeV) & $m_\sigma$ (MeV) & $m_\zeta$ (MeV) \\
\hline
0 & 0.16 & -16.0 & 0.742 & 303 & 220.2 & 487.8 & 1168.0  \\
0.6 & 0.16 & -16.0 & 0.731 & 359 & 215.5 & 490.0 & 1173.4  \\
0.75 & 0.16 & -16.0 & 0.713 & 473 & 209.5 & 491.6 & 1181.2  \\
\end{tabular}
\end{center}
\end{table}

\bigskip
\bigskip

\begin{center}
\begin{figure}[hbt]
\includegraphics[scale=0.66]{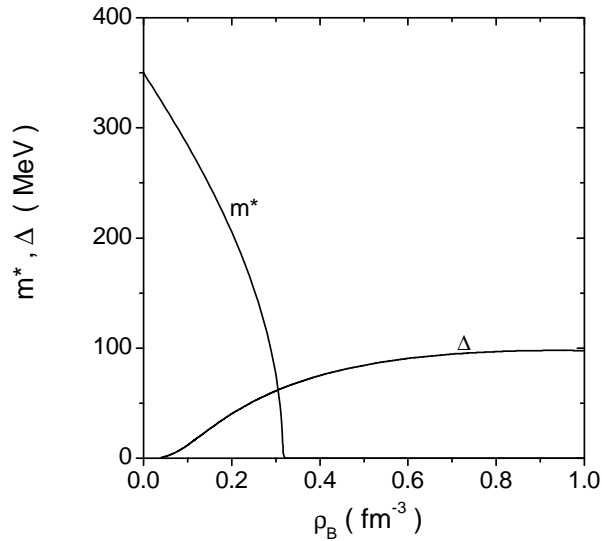}
\caption{Effective quark mass and superconducting energy gap calculated
in the NJL model versus baryon density at zero temperature.}
\end{figure}

\begin{figure}[hbt]
\includegraphics[scale=0.66]{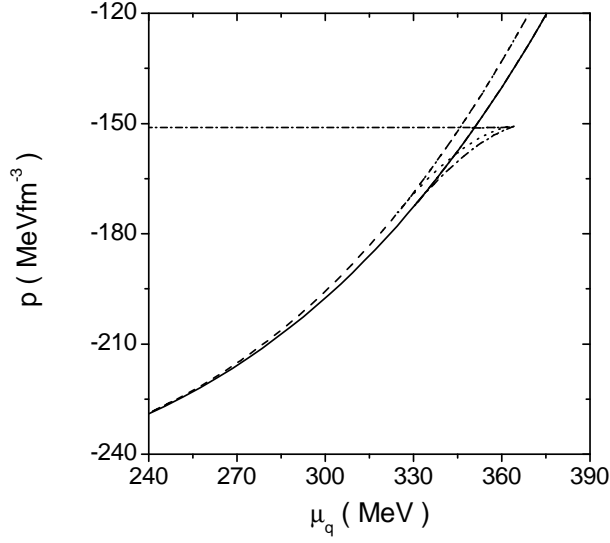}
\caption{Thermodynamic potential of quark matter versus quark
chemical potential, $\mu_q$, at zero temperature. The solid, dashed,
dotted and dash-dotted lines are for $m^*=0$, $\Delta=0$, $m^*=0$,
$\Delta\neq 0$, $m^*\neq 0$, $\Delta\neq 0$ and $m^*\neq 0$,
$\Delta=0$, respectively.}
\end{figure}

\begin{figure}[hbt]
\includegraphics[scale=0.66]{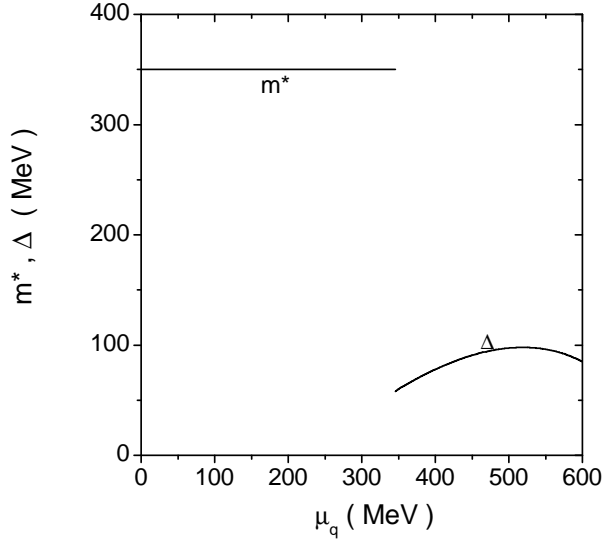}
\caption{Effective quark mass and superconducting energy gap versus
quark chemical potential at zero temperature.}
\end{figure}

\begin{figure}[hbt]
\includegraphics[scale=0.66]{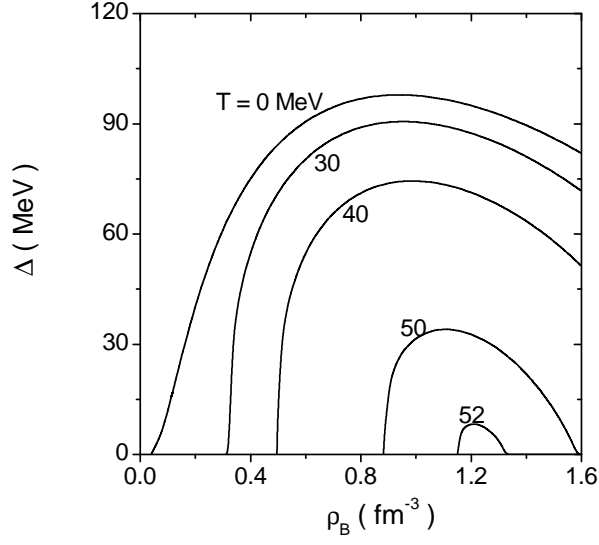}
\caption{Superconducting energy gap versus baryon density at
different temperatures.}
\end{figure}

\begin{figure}[hbt]
\includegraphics[scale=0.66]{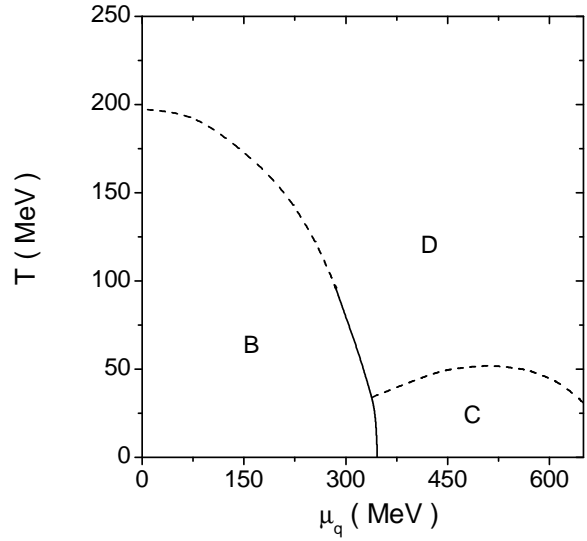}
\caption{T-$\mu$ phase diagram for quark matter. The solid line
stands for the first order phase transition and dashed lines stand
for the second order phase transition. B, C and D correspond to the
different phases explained in the text.}
\end{figure}

\begin{figure}[hbt]
\includegraphics[scale=0.66]{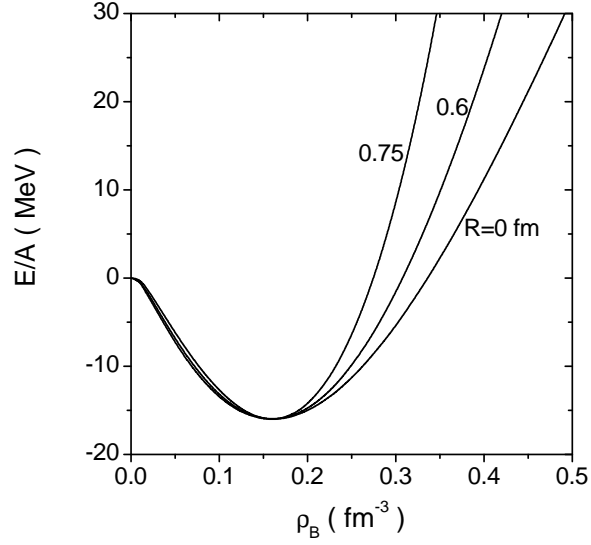}
\caption{The energy per baryon of nuclear matter versus baryon density.
The radius of the nucleon is chosen to be 0, 0.6 fm and 0.75 fm, respectively.}
\end{figure}

\begin{figure}[hbt]
\includegraphics[scale=0.66]{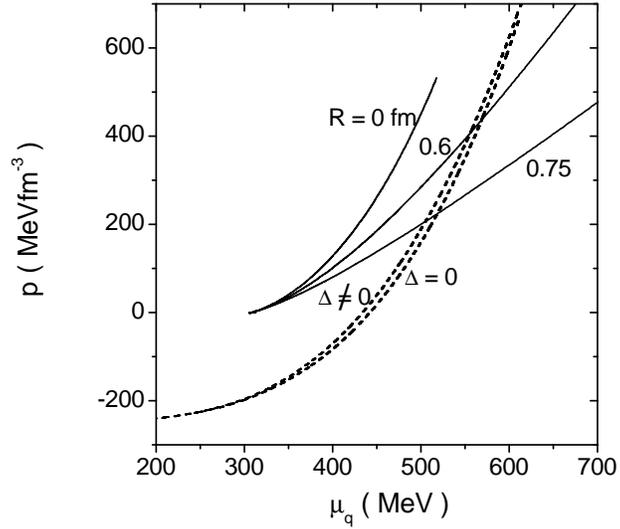}
\caption{Pressure of nuclear matter and quark matter versus
chemical potential. The solid lines are for nuclear matter and dashed lines
are for quark matter.}
\end{figure}

\begin{figure}[hbt]
\includegraphics[scale=0.66]{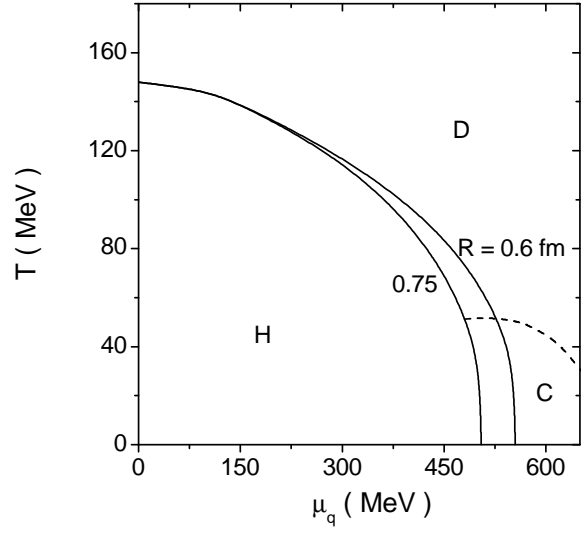}
\caption{T-$\mu$ phase diagram for nuclear matter and quark matter.
The solid lines are for first order phase transition between nuclear
matter and quark matter. The dashed line is for the second order
phase transition between superconducting quark matter and normal
quark matter. H is for the phase of nuclear matter and C and D are for
the quark phases.}
\end{figure}

\begin{figure}[hbt]
\includegraphics[scale=0.66]{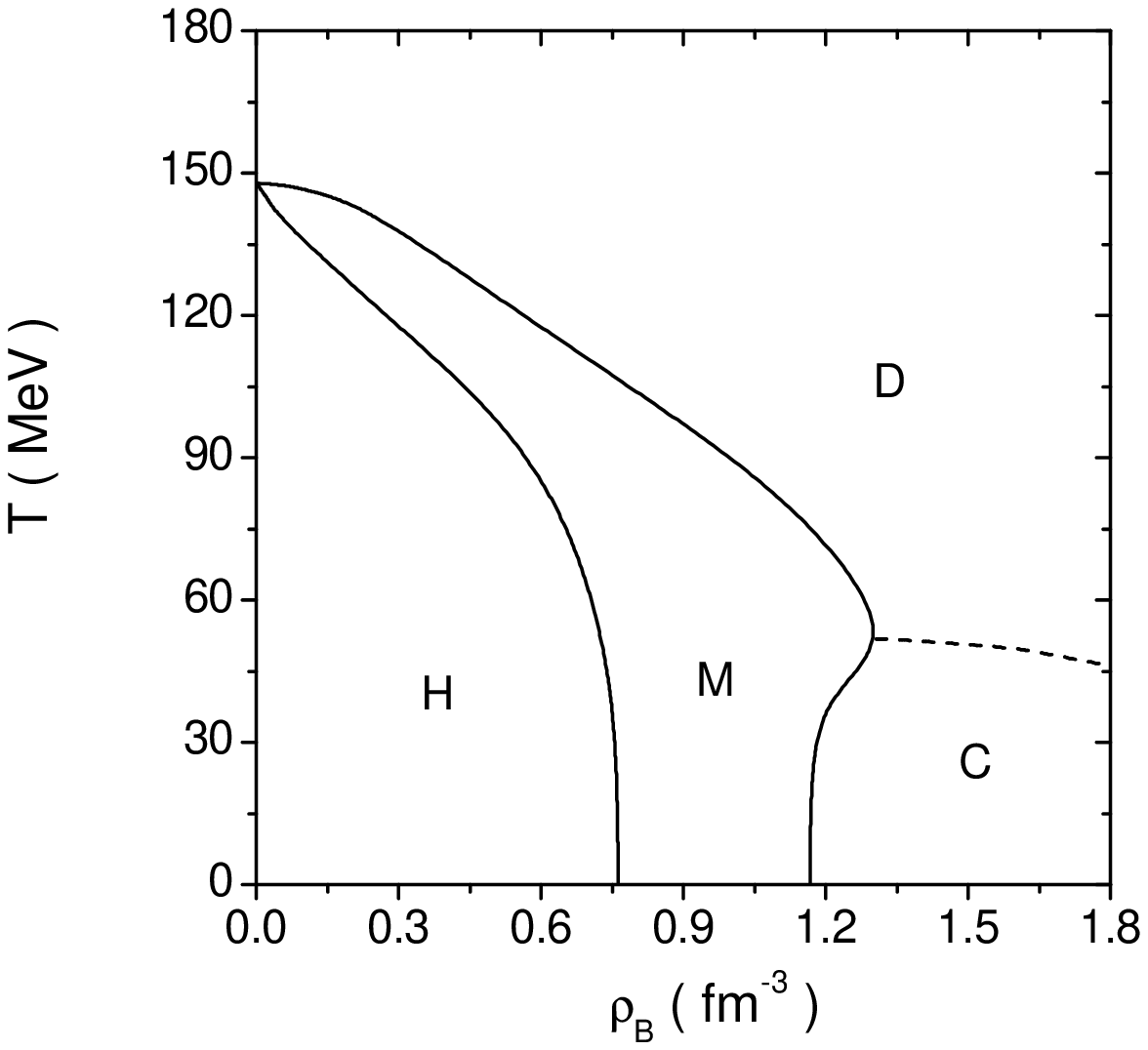}
\caption{T-$\rho$ phase diagram for nuclear matter and quark matter.
The radius of nucleon is chosen to be 0.6 fm. H is for the phase of
nuclear matter. M is for the mixed phase. C and D are for the quark
phases.}
\end{figure}

\begin{figure}[hbt]
\includegraphics[scale=0.66]{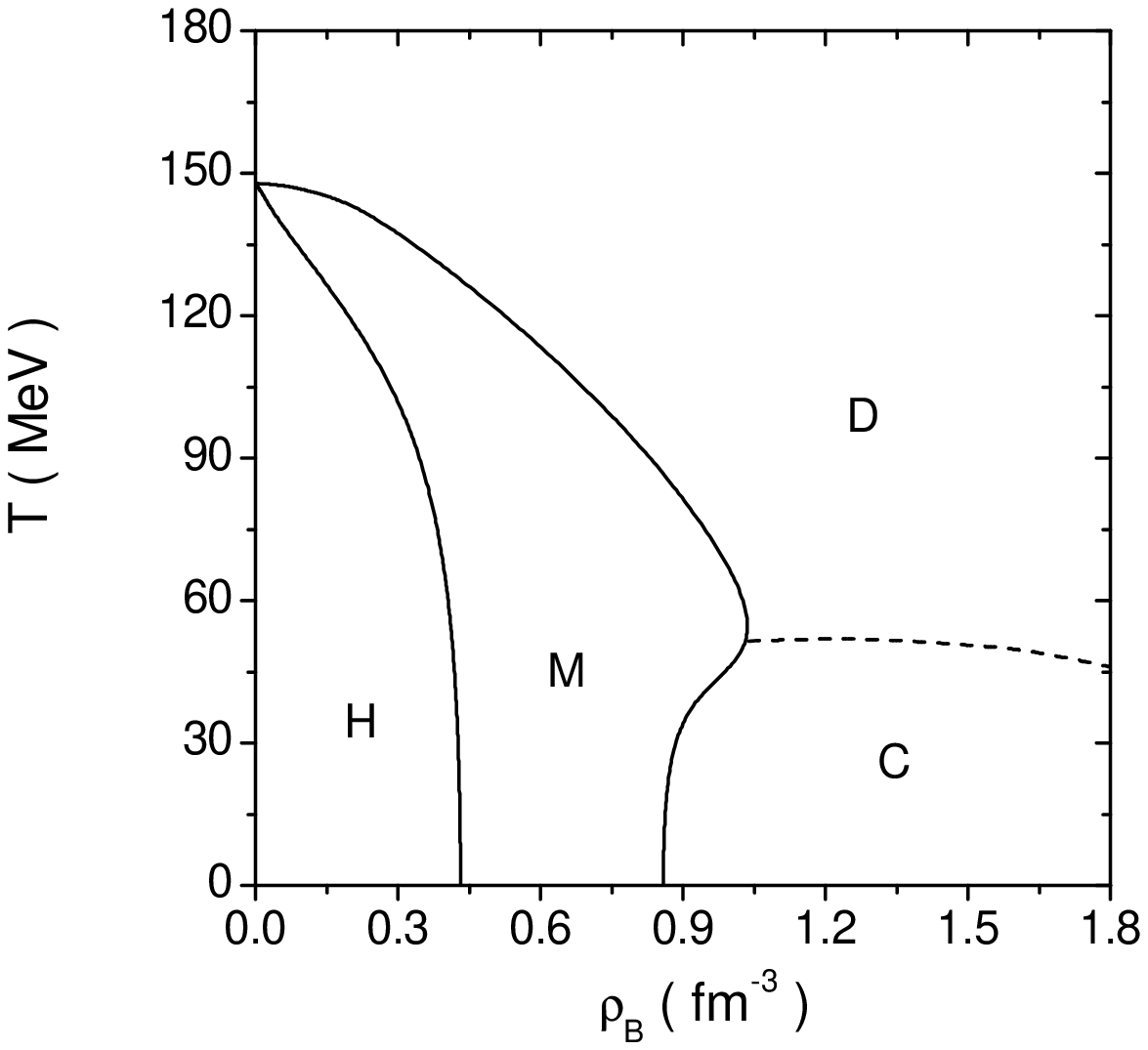}
\caption{T-$\rho$ phase diagram for nuclear matter and quark matter.
The radius of the nucleon is chosen to be 0.75 fm. H is for the phase of
nuclear matter. M is for the mixed phase. C and D are for the quark
phases.}
\end{figure}

\begin{figure}[hbt]
\includegraphics[scale=0.66]{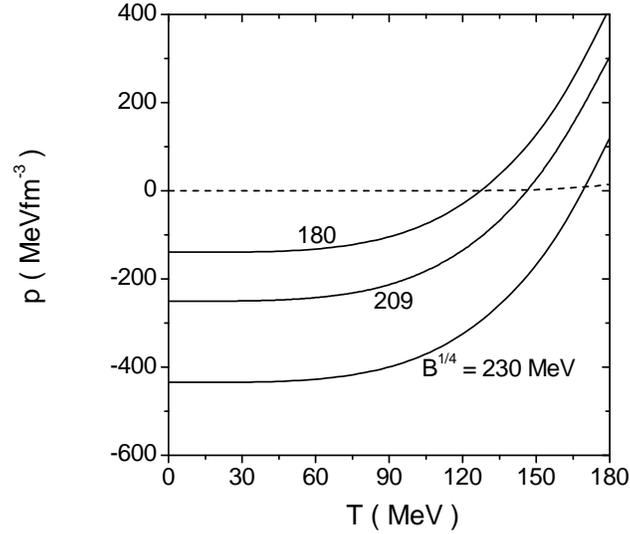}
\caption{Pressure of nuclear and quark matter at zero baryon density
versus temperature. The solid lines are for quark matter with the
different bag constants. The dashed line is for nuclear matter.}
\end{figure}
\end{center}

\end{document}